\documentstyle[preprint,aps,epsf]{revtex}
\begin{document}
\draft
\title{
The rare decay $\eta \rightarrow \pi \pi \gamma \gamma $ in chiral 
perturbation theory 
\thanks{Supported by the Deutsche Forschungsgemeinschaft 
(SFB 201)}}
\author{G. Kn\"{o}chlein\footnote{Present 
address: Department of Physics and Astronomy, University 
of Massachusetts, Amherst, MA 01003}, S. Scherer, D. Drechsel}
\address{
Institut f\"{u}r Kernphysik, 
Johannes Gutenberg-Universit\"{a}t, D-55099 Mainz, Germany}
\date{\today}
\maketitle

\begin{abstract}
We investigate the rare radiative eta decay modes 
$\eta \rightarrow \pi^+ \pi^- \gamma \gamma$ 
and 
$\eta \rightarrow \pi^0 \pi^0 \gamma \gamma$ 
within the framework of chiral perturbation theory at ${\cal{O}} (p^4)$. 
We present photon spectra and partial decay rates for both processes 
as well as a Dalitz contour plot for the charged decay. 
\end{abstract}
\newpage
\narrowtext
\section{Introduction}
With the commissioning of new powerful facilities 
with high production rates, low- and medium-energy meson physics will 
experience renewed interest.
$B$ physics at higher energies and $K$ physics at 
lower energies have stimulated large efforts on the 
experimental as well as the theoretical side; 
the field of $\eta$ and $\eta'$ physics 
provides a considerable amount of open questions, and the new facilities are
expected to address them from the experimental side. 
The anticipated numbers of $10^{8}$ - $10^9$ observed etas per year at 
CELSIUS ($\sim 2.2 \cdot 10^9$), 
ITEP ($\sim (0.27 - 2.7)\cdot 10^9$) and DA$\Phi$NE 
($\sim 3.2 \cdot 10^8$) \cite{DAFNE1} 
will allow for experiments which on the one hand 
supply precise figures on the more frequent eta decays and which on the other
hand focus on rare eta decays. 
Such rare decays can supply valuable information on anomalous processes or a 
possible $C$ violation in eta decays. 
Another interesting question is whether it will be feasible to observe some 
of the rare eta decays at the new laser backscattering facility GRAAL at 
Grenoble ($\sim 10$ decay events per second)
or even at the $cw$ electron facilities at Mainz (MAMI), 
Bonn (ELSA) and Newport News (CEBAF), where 
many eta photo- and electroproduction experiments are scheduled or already 
being carried through. 
On the 
theoretical side the development of chiral perturbation theory 
\cite{Weinberg,GL} as an 
effective theory for the confinement phase of QCD has supplied a 
consistent framework for the calculation of low-energy processes in the 
$SU\!\left(2\right)$ as well as the $SU\!\left(3\right)$ flavor 
sector of QCD. Chiral
perturbation theory has turned out to be a valuable tool in the investigation 
of meson interactions at low energies. However, 
the theory seems to work much better in the $SU\!\left(2\right)$ sector 
than in the 
$SU\!\left(3\right)$ sector due to the comparatively large mass of the 
strange quark. In particular, processes involving the eta meson have 
confronted chiral perturbation theory with various problems 
which could only be solved partly by considering next-to-leading-order terms 
and electromagnetic corrections or by going even beyond next-to-leading order 
(see, e.g., \cite{GLe,Pich}). 
These problems are also related to 
the large \mbox{$\eta$ - $\eta'$} mixing angle and the 
$U_A(1)$ problem. 
In pure chiral perturbation theory up to ${\cal{O}}(p^4)$, 
the eta singlet field is integrated out \cite{GL}, 
because the mass of the corresponding physical $\eta'$, $m_{\eta'}=
957.7\,{\mathrm{MeV}}$, 
is larger than the low-energy scale 
$\Lambda_{CPT} \approx m_{\rho}$. 
On the other hand, current-algebra-like calculations with phenomenologically 
determined decay constants and $\eta$ - $\eta'$ mixing, which keep the eta 
singlet field as an explicit degree of freedom, 
yield reasonable results. 
Since we will restrict our calculation of the process 
$\eta \rightarrow \pi\pi\gamma\gamma$ to tree level,
we will choose such a phenomenologically inspired approach. 
The charged mode of this 
process is of particular interest, because it can supply information on a
Wess-Zumino-Witten contact term \cite{WZ,Witten} 
involving the interaction of 
three mesons and two photons (for a review of anomalous processes see, e.g.,
\cite{bijnensrev}). 
The eta meson is the only particle of the pseudoscalar octet which decays 
through such a mechanism. 
Another possibility of testing this special kind of contact
term in a scattering experiment is the process $\gamma\gamma\rightarrow
\pi^+\pi^-\pi^0$ which has recently 
been treated in chiral perturbation theory 
\cite{Bos}. Due to the lack of meson collision facilities, it
is difficult to extract information on this vertex from a scattering experiment
with a meson in the initial state, because in practice such experiments 
involve virtual particles and necessitate uncertain
extrapolations. 
On the experimental side a data analysis dating back to 
the sixties with rather low
statistics gives 
upper bounds on the branching ratios of the charged 
mode $\eta \rightarrow \pi^+\pi^-\gamma\gamma$ \cite{PDG94,PC,bal}: 
\begin{eqnarray}
\protect{\cite{PC}}:& 
\Gamma\left(\eta\rightarrow
\pi^+\pi^-\gamma\gamma\right)/\Gamma_{tot} & < 2.1\times 10^{-3}, 
\label{e1} \\
\protect{\cite{bal}}:& \Gamma\left(\eta\rightarrow
\pi^+\pi^-\gamma\gamma\right)/\Gamma_{tot} & < 3.7\times 10^{-3}.
\label{e2}
\end{eqnarray}
In the analyses \cite{PC} and \cite{bal} the upper limits for the branching
ratios were derived for a missing mass of neutral particles larger than 
$195 \, \mathrm{MeV}$. Hopefully, the situation will improve when future
experiments will be carried through.  

\section{Kinematics and Observables}
The four-momenta and polarization vectors for the charged
decay mode $\eta\rightarrow\pi^+\pi^-\gamma\gamma$ are defined in Fig.~1. 
For the neutral decay mode we use analogous
descriptors.  
The full kinematics of a decay process with four particles in the final 
state requires five independent kinematical variables (see, e.g., 
\cite{CaMa,DAFNE}). 
For the definition of these variables we 
will consider three reference frames: the rest system of the eta meson 
$\Sigma_{\eta}$, the dipion center-of-mass system $\Sigma_{\pi\pi}$, and the 
diphoton center-of-mass system $\Sigma_{\gamma\gamma}$. Our kinematical 
variables are (see Fig.~2)
\begin{itemize}
\item $s_{\pi}$, the square of the center-of-mass energy of the pions,
\item $s_{\gamma}$, the square of the center-of-mass energy of the photons,
\item $\theta_{\pi_1}$, the angle of the pion with momentum $k_1$ in 
$\Sigma_{\pi\pi}$ with respect to the direction of flight of the dipion in 
$\Sigma_{\eta}$,
\item $\theta_{\gamma_1}$, the angle of the photon with momentum $q_1$ in 
$\Sigma_{\gamma\gamma}$ with respect to the direction of flight of the 
diphoton in 
$\Sigma_{\eta}$, 
\item $\phi$, the angle between the plane formed by the pions in 
$\Sigma_{\eta}$ and the corresponding plane formed by the photons.
\end{itemize}
In order to define these variables more precisely we 
introduce a unit vector ${\hat{v}}$ along the direction of flight of the 
dipion in $\Sigma_{\eta}$, and unit vectors ${\hat{c}}$ and 
${\hat{d}}$ along 
the projections of $\vec k_1$ perpendicular to ${\hat{v}}$ and 
of $\vec q_1$ perpendicular to $-{\hat{v}}$, respectively,
\begin{eqnarray}
{\hat{c}} & = & \left( \vec k_1 - {\hat{v}} {\hat{v}} \cdot \vec k_1 \right)
/ \left[\vec k_1^{\,2} - \left(\vec k_1 \cdot {\hat{v}} \right)^2 
\right]^{1/2},\\
{\hat{d}} & = & \left( \vec q_1 - {\hat{v}} {\hat{v}} \cdot \vec q_1 \right)
/ \left[\vec q_1^{\,2} - \left(\vec q_1 \cdot {\hat{v}} \right)^2 
\right]^{1/2}.
\end{eqnarray}
With these definitions the five kinematical variables are defined as 
follows:
\begin{eqnarray}
s_{\pi} & = & \left(k_1+k_2\right)^2,\\
s_{\gamma} & = & \left(q_1+q_2\right)^2,\\
\cos \theta_{\pi_1} & = & {\hat{v}} \cdot \vec k_1 / \mid\! \vec k_1 \!\mid,\\
\cos \theta_{\gamma_1} & = & - {\hat{v}} \cdot \vec q_1 / \mid
\! \vec q_1 \!\mid,\\
\cos \phi & = & {\hat{c}} \cdot {\hat{d}}.
\end{eqnarray}
The physical region of the decay process is reflected in the range of the 
kinematical variables:
\begin{eqnarray}
0 \leq & s_{\gamma} & \leq \left( m_{\eta} - 2 m_{\pi} \right)^2,\\
4 m_{\pi}^2 \leq & s_{\pi} & \leq \left( m_{\eta} - \sqrt{s_{\gamma}}
\right)^2,\\
0 \leq & \theta_{\pi_1} & \leq \pi,\\
0 \leq & \theta_{\gamma_1} & \leq \pi,\\
0 \leq & \phi & \leq 2 \pi.
\end{eqnarray}
The invariant matrix element squared, 
$\mid\!{\cal{\overline{M}}}\!\mid^2$, will be expressed in terms of  
Lorentz scalar products of the five momenta $k_1$,$k_2$,$q_1$, 
$q_2$ and $p$. 
One of these momentum vectors can be eliminated because of momentum
conservation. In order to express the Lorentz scalar products in terms of the 
kinematical variables specified above we now introduce adequate 
linear combinations of the momenta:
\begin{eqnarray}
K & = & k_1 + k_2,\\
L & = & k_1 - k_2,\\
Q & = & q_1 + q_2,\\
R & = & q_1 - q_2.
\end{eqnarray} 
For further reference we need the expressions
\begin{eqnarray}
K \cdot K & = & s_{\pi},\\
Q \cdot Q & = & s_{\gamma},\\
K \cdot Q & = & \frac{1}{2} \left( m_{\eta}^2 - s_{\pi} - s_{\gamma} \right)
,\\
K \cdot R & = & x \cos \theta_{\gamma_1},\\
L \cdot Q & = & \sigma_{\pi} x \cos \theta_{\pi_1}, \\
L \cdot R & = & \sigma_{\pi} \left[ K \cdot Q \cos \theta_{\pi_1} 
\cos \theta_{\gamma_1} \right.\nonumber \\
& & \left.- \left( s_{\pi} s_{\gamma} \right)^{1/2} 
\sin \theta_{\pi_1} \sin \theta_{\gamma_1} \cos \phi \right],
\end{eqnarray}
with
\begin{eqnarray}
x & = & \sqrt{\left( K \cdot Q \right)^2 - s_{\pi} s_{\gamma}},\\
\sigma_{\pi} & = & \sqrt{1 - 4 m_{\pi}^2/s_{\pi}}.
\end{eqnarray}
The ten Lorentz scalar products in 
$\mid \!{\overline{\cal{M}}}\!\mid^2$ can now be expressed as follows:
\begin{eqnarray}
k_1 \cdot k_2 & = & \frac{1}{2} \left( K \cdot K - 2 m_{\pi}^2\right),\\
q_1 \cdot q_2 & = & \frac{1}{2} Q \cdot Q,\\
k_1 \cdot q_1 & = & \frac{1}{4} \left(K \cdot Q + L \cdot Q + K \cdot R + L 
\cdot R \right),\\
k_2 \cdot q_1 & = & \frac{1}{4} \left(K \cdot Q - L \cdot Q + K \cdot R - L 
\cdot R \right),\\
k_1 \cdot q_2 & = & \frac{1}{4} \left(K \cdot Q + L \cdot Q - K \cdot R - L 
\cdot R \right),\\
k_2 \cdot q_2 & = & \frac{1}{4} \left(K \cdot Q - L \cdot Q - K \cdot R + L 
\cdot R \right),\\
p \cdot k_1 & = & k_1 \cdot q_1 + k_1 \cdot q_2 +k_1 \cdot k_2 + m_{\pi}^2,\\
p \cdot k_2 & = & k_2 \cdot q_1 + k_2 \cdot q_2 +k_1 \cdot k_2 + m_{\pi}^2,\\
p \cdot q_1 & = & q_1 \cdot q_2 + k_1 \cdot q_1 + k_2 \cdot q_1,\\
p \cdot q_2 & = & q_1 \cdot q_2 + k_1 \cdot q_2 + k_2 \cdot q_2.
\end{eqnarray}
The differential decay rate can be written as
\begin{eqnarray}
\label{width}
{\mathrm{d}}^5\Gamma \left(\eta \rightarrow \pi\pi\gamma\gamma\right) 
& = &
2^{-14} \pi^{-6} m_{\eta}^{-3} C^{-1} \sigma_{\pi} x 
\mid\! {\overline{\cal{M}}} 
\!\mid^2 
\nonumber \\
& & {\mathrm{d}}s_{\pi} {\mathrm{d}}s_{\gamma} 
{\mathrm{d}}\!\cos\theta_{\pi_1} 
{\mathrm{d}}\!\cos\theta_{\gamma_1} {\mathrm{d}}\phi,
\end{eqnarray}
where the symmetry factor $C$ is equal to $4$ in the decay 
$\eta \rightarrow \pi^0\pi^0\gamma\gamma$ because of two pairs of identical 
particles in the final state, and equal to $2$ in the decay 
$\eta \rightarrow \pi^+\pi^-\gamma\gamma$
because of one identical particle pair in the final state. Now we will proceed
to investigate how chiral dynamics manifests itself in the Lorentz-invariant
matrix element ${\cal{M}}$.

\section{Chiral Dynamics of the decay $\eta \rightarrow \pi\pi\gamma\gamma$}
We will restrict our calculation of ${\cal{M}}$ to the leading-order 
contributions. 
Since the process $\eta \rightarrow \pi\pi\gamma\gamma$ 
involves the electromagnetic interaction of an odd number of pseudoscalar
mesons, 
the leading contributions must contain a vertex of odd intrinsic parity. Such a
vertex is at least of ${\cal{O}}\!\left(p^4\right)$ in the momentum expansion,
and thus, according to Weinberg's power counting \cite{Weinberg}, 
we expect the leading
contribution to be of ${\cal{O}}\!\left(p^4\right)$. 
Consequently, the 
interaction Lagrangian we will use for our tree-level calculation 
contains the standard 
${\cal{O}}\!\left(p^2\right)$ piece \cite{Weinberg} 
and the anomalous Wess-Zumino-Witten Lagrangian \cite{WZ,Witten}, 
but no terms from the Gasser-Leutwyler Lagrangian \cite{GL} of 
${\cal{O}}\!\left(p^4\right)$:
\widetext
\begin{eqnarray}
{\cal{L}} & = & {\cal{L}}^{(2)}+{\cal{L}}^{(4)}_{WZW} \nonumber \\
& = &  
\frac{F_{\pi}^2}{4} {\mathrm{tr}} ((D^{\mu} U)^{\dagger} D_{\mu} U)
+ \frac{F_{\pi}^2}{4} {\mathrm{tr}}(\chi^{\dagger} U + \chi U^{\dagger})
\nonumber\\
&&
+\frac{eN_c}{48\pi^2}\varepsilon^{\mu\nu\alpha\beta}A_\mu {\mathrm{tr}}
(Q\partial_\nu U \partial_\alpha U^\dagger\partial_\beta U U^\dagger
-Q\partial_\nu U^\dagger\partial_\alpha U\partial_\beta U^\dagger U)
\nonumber\\ 
&& 
- \frac{i e^2 N_c}{24 \pi^2} \varepsilon^{\mu\nu\alpha\beta}\partial_{\mu}
A_{\nu} A_{\alpha} {\mathrm{tr}}(Q^2(U \partial_{\beta} U^{\dagger} 
+ \partial_{\beta} U^{\dagger} U )
-\frac{1}{2}Q U^{\dagger} Q \partial_{\beta} U + \frac{1}{2} Q U Q
\partial_{\beta} U^{\dagger}), \label{genlag}
\end{eqnarray}
\narrowtext
where $\varepsilon_{0123}=1$.
In Eq.\ (\ref{genlag}) we have only listed those terms of 
${\cal{L}}^{(4)}_{WZW}$ 
which actually give a contribution to the invariant amplitude. 
The covariant derivative is defined as 
\begin{equation}
D_{\mu} U =  
\partial_{\mu} U + i e
A_{\mu} \left[ Q, U \right], 
\end{equation}
where the matrix $Q$ represents the electromagnetic charges 
of the three flavors 
in $SU\!\left(3\right)$, 
\begin{equation}
Q={\mathrm{diag}} \left(2,-1,-1 \right)/3.
\end{equation}
The matrix
\begin{equation}
\chi = 2 B_0 m
\end{equation}
contains the quark masses, 
\begin{equation}
m = {\mathrm{diag}}\left(m_u,m_d,m_s\right),
\end{equation} 
where $B_0$ is related to the quark condensate and is
given by the relation $\left(m_u+m_d\right) B_0 = m_{\pi}^2$. 
The meson field operators are represented by the matrix 
$U = \mathrm{exp}\left(i \Phi/F_{\Phi} \right)$, where 
the nonet field matrix $\Phi$ can be decomposed into an octet and a singlet 
part, $\Phi = \Phi_8 + \Phi_1$, with
\begin{equation}
\Phi_8 =  
\left(
\begin{array}{ccc}
\pi_3 + \frac{1}{\sqrt{3}} \eta_8 & \sqrt{2} \pi^+ & 0 \\
\sqrt{2} \pi^- & - \pi_3 + \frac{1}{\sqrt{3}} \eta_8 & 0 \\
0 & 0 & - \frac{2}{\sqrt{3}} \eta_8 
\end{array}
\right)
\end{equation}
and
\begin{equation}
\Phi_1 = \sqrt{\frac{2}{3}} \eta_0 \,
{\mathrm{diag}} \left( 1, 1, 1 \right).
\end{equation}
In the expansion of $U$ the decay constants 
$F_{\pi}$, $F_8$ or ${\overline{F}}_0$ will be inserted 
for the constant $F_{\Phi}$, depending on whether the constant belongs to a 
$\pi$, $\eta_8$ or $\eta_0$ field. We will use $F_{\pi}=93\,{\mathrm{MeV}}$, 
$F_8=1.25 F_{\pi}$ and ${\overline{F}}_0=1.06 F_{\pi}$  
as numerical values \cite{GK}. 
Our calculation in chiral perturbation theory will be carried out with the 
group theoretical octet and singlet eta states, $\mid\!\eta_8 \rangle$ 
and $\mid\!\eta_0 \rangle$. We will introduce $\eta$-$\eta'$ mixing via the 
phenomenological mixing angle $\theta = -20^\circ$ \cite{GK}:  
\begin{eqnarray}
|\eta\rangle & = & \cos \theta |\eta_8 \rangle - \sin \theta | \eta_0 \rangle
, \\ 
|\eta' \rangle & = & \sin \theta |\eta_8 \rangle + \cos \theta |\eta_0 \rangle.
\end{eqnarray} 
The Feynman diagrams contributing to the charged decay mode 
$\eta\rightarrow\pi^+\pi^-\gamma\gamma$ 
of 
${\cal{O}}\!\left(p^4\right)$ are 
displayed in Fig.~\ref{feyndiagcharged}. 
There are three different classes of 
diagrams at tree level: diagrams with a four-meson vertex of 
${\cal{O}}\!\left(p^2\right)$, a propagating 
neutral meson, and a decay vertex into two photons of 
${\cal{O}}\!\left(p^4\right)$ 
(class 1), 
Wess-Zumino-Witten contact terms 
of ${\cal{O}}\!\left(p^4\right)$ 
(class 2) and 
internal bremsstrahlung diagrams, where one photon is emitted off a charged 
pion line (classes 3.1 and 3.2). 
The first class of diagrams is gauge invariant 
by itself, whereas the amplitudes corresponding to the diagrams from the 
second and third class have to be added in order to obtain gauge 
invariance. In the neutral decay mode $\eta\rightarrow\pi^0\pi^0\gamma\gamma$ 
only the first class 
of diagrams is relevant (Fig.~\ref{feyndiagneutral}). 

Starting from the general chiral Lagrangian of Eq.\ (\ref{genlag}), we now 
list the interaction terms 
relevant for the process $\eta \rightarrow 
\pi^+\pi^-\gamma\gamma$:
\begin{eqnarray}
{\cal{L}}^{(4),2\gamma3\phi}_{WZW} & = & 
\frac{e^2 N_c}{12 \sqrt{3} \pi^2 F_{\pi}^2}
\varepsilon^{\mu\nu\alpha\beta}
\partial_{\mu}A_{\nu} A_{\alpha} 
\pi^+ \pi^-
\nonumber \\
& & 
\times
\left(\frac{1}{F_8} \partial_{\beta} \eta_8 + 
\frac{\sqrt{2}}{{\overline{F}}}_0
\partial_{\beta} \eta_0
\right), \label{lanf} \label{li} \\
{\cal{L}}^{(4),1\gamma3\phi}_{WZW} & = & 
\frac{i e N_c}{12 \sqrt{3} \pi^2 F_{\pi}^2}
\varepsilon^{\mu\nu\alpha\beta}
A_{\mu}
\partial_{\nu}\pi^+\partial_{\alpha}\pi^-
\nonumber \\
& & 
\times
\left(\frac{1}{F_8} \partial_{\beta} \eta_8 + 
\frac{\sqrt{2}}{{\overline{F}}}_0
\partial_{\beta} \eta_0
\right), \\
{\cal{L}}^{(4),2\gamma1\phi}_{WZW} & = & 
-\frac{ e^2 N_c}{24 \pi^2 }
\varepsilon^{\mu\nu\alpha\beta}
\partial_{\mu} A_{\nu} A_{\alpha} 
\nonumber \\
& & 
\times
\partial_{\beta}\left(
\frac{1}{F_{\pi}}\pi^0+
\frac{1}{\sqrt{3}F_8} \eta_8 + \frac{2\sqrt{2}}{\sqrt{3}
\,{\overline{F}}_0}
\eta_0
\right), \\
{\cal{L}}^{(2),4\phi} & = & 
\frac{ F_{\pi}^2 B_0}{24}
\left(
\frac{4 \left(m_u+m_d\right)}{F_{\pi}^2F_8^2}\eta_8\eta_8\pi^+\pi^-
\right. \nonumber \\
& & 
\left.
+
\frac{8 \sqrt{2} \left(m_u+m_d\right)}{F_{\pi}^2 F_8 {\overline{F}}_0}
\eta_8\eta_0\pi^+\pi^-\right.\nonumber\\
& & \left.+
\frac{8 \left(m_u+m_d\right)}{F_{\pi}^2 {\overline{F}}_0^2}
\eta_0\eta_0\pi^+\pi^-
+
\right. \nonumber \\
& & 
\left.
\frac{8 \left(m_u-m_d\right)}{\sqrt{3}F_{\pi}^3 F_8}
\eta_8\pi^+\pi^-\pi^0
\right. \nonumber \\
& & 
\left.
+
\frac{8 \sqrt{2}\left(m_u-m_d\right)}{\sqrt{3}F_{\pi}^3 {\overline{F}}_0}
\eta_0\pi^+\pi^-\pi^0
\right), \label{vpvc} \\
{\cal{L}}^{(2),1\gamma 2\phi} & = & 
i e A^{\rho}\left(\partial_{\rho}\pi^-\pi^+-\partial_{\rho}\pi^+\pi^-\right).
\label{lend}
\end{eqnarray}
The invariant amplitude for $\eta\rightarrow\pi^+\pi^-\gamma\gamma$ is then
given as the sum of the amplitudes from the three classes of Feynman
diagrams (Fig.~\ref{feyndiagcharged}),
\begin{equation}
{\cal{M}}={\cal{M}}_1+{\cal{M}}_2+{\cal{M}}_3,
\end{equation}
where
\widetext
\begin{eqnarray}
{\cal{M}}_1 & = & - \frac{i e^2 N_c}{12 \sqrt{3} \pi^2} 
\varepsilon^{\mu\nu\alpha\beta} \varepsilon_{1,\mu}
\varepsilon_{2,\alpha} q_{1,\nu} q_{2,\beta} 
\left\{
\frac{B_0\left(m_u+m_d\right)}{3 \left( 2 q_1 \cdot q_2 - m_{\eta_8}^2 
\right)}
\left(
\frac{\cos \theta}{F_8^3}
-\frac{\sqrt{2}\sin \theta}{F_8^2 {\overline{F}}_0}
\right)
\right.\nonumber\\
& &  +
\frac{4 B_0\left(m_u+m_d\right)}{3 \left( 2 q_1 \cdot q_2 - m_{\eta_0}^2
\right)}
\left(
\frac{\cos \theta}{{\overline{F}}_0^2 F_8}
-\frac{\sqrt{2}\sin \theta}{{\overline{F}}_0^3}
\right)
\nonumber\\
&&\left. +
\frac{B_0\left(m_u-m_d\right)}{3 \left( 2 q_1 \cdot q_2 - m_{\pi_3}^2
\right)}
\left(
\frac{\cos \theta}{F_{\pi}^2 F_8}
-\frac{\sqrt{2}\sin \theta}{F_{\pi}^2 {\overline{F}}_0}
\right)
\right\},\\
{\cal{M}}_2 & = &
-\frac{i e^2 N_c}{12 \sqrt{3} \pi^2} 
\varepsilon^{\mu\nu\alpha\beta}
p_{\beta} \varepsilon_{1,\mu} \varepsilon_{2,\alpha} \left(q_1 - q_2 
\right)_{\nu}
\left(
\frac{\cos \theta}{F_{\pi}^2 F_8}
-\frac{\sqrt{2}\sin \theta}{F_{\pi}^2 {\overline{F}}_0}
\right),\\
{\cal{M}}_3 & = &
\frac{i e^2 N_c}{12 \sqrt{3} \pi^2} 
\varepsilon^{\mu\nu\alpha\beta}
\left(
\frac{\cos \theta}{F_{\pi}^2 F_8}
-\frac{\sqrt{2}\sin \theta}{F_{\pi}^2 {\overline{F}}_0}
\right)
\left\{
\varepsilon_{1,\mu}
q_{1,\alpha} k_{2,\nu} p_{\beta} 
\frac{\varepsilon_2 \cdot k_1}{q_2 \cdot k_1}
+
\varepsilon_{2,\mu}
q_{2,\alpha} k_{2,\nu} p_{\beta} 
\frac{\varepsilon_1 \cdot k_1}{q_1 \cdot k_1}
\right.\nonumber \\
& & \left. 
+
\varepsilon_{1,\mu}
q_{1,\alpha} k_{1,\nu} p_{\beta} 
\frac{\varepsilon_2 \cdot k_2}{q_2 \cdot k_2}
+
\varepsilon_{2,\mu}
q_{2,\alpha} k_{1,\nu} p_{\beta} 
\frac{\varepsilon_1 \cdot k_2}{q_1 \cdot k_2}
\right\}.
\end{eqnarray}
\narrowtext
The masses in the propagators can be expressed in terms of the physical 
masses using the relations 
\begin{eqnarray}
m_{\eta_8}^2 & = & m_{\eta}^2 \cos^2 \theta + m_{\eta'}^2 \sin^2 \theta, \\
m_{\eta_0}^2 & = & m_{\eta}^2 \sin^2 \theta + m_{\eta'}^2 \cos^2 \theta, \\
m_{\pi_3}^2 & = & m_{\pi^0}^2. 
\end{eqnarray}
As we will not be concerned with
photon polarizations in the final state, 
we carry out the sum over the polarizations of
the real photons, 
\begin{equation}
\mid\! {\overline{\cal{M}}} \!\mid^2
=
\sum_{\kappa_1,\kappa_2} 
{\cal{M}}\left(\kappa_1,\kappa_2\right)
{\cal{M}}^*\left(\kappa_1,\kappa_2\right).
\end{equation}
For that purpose we exploit the completeness 
relation 
\begin{equation}
\sum_{\kappa} \varepsilon_{\gamma}\left(\kappa\right) 
\varepsilon_{\delta}^* \left(\kappa\right) \rightarrow - g_{\gamma\delta},
\end{equation}
which is based on current conservation (see, e.g., \cite{HM}): 
\widetext
\begin{eqnarray}
\sum_{\kappa_1,\kappa_2}
{\cal{M}}_i\left(\kappa_1, \kappa_2\right)
{\cal{M}}^*_j\left(\kappa_1,\kappa_2\right) 
& = & \sum_{\kappa_1,\kappa_2} 
{\cal{M}}_i^{\lambda\sigma}
{{\cal{M}}_j^{\rho\tau}}^*
\varepsilon_{1,\lambda}\left(\kappa_1\right)
\varepsilon_{1,\rho}^*\left(\kappa_1\right)
\varepsilon_{2,\sigma}\left(\kappa_2\right)
\varepsilon_{2,\tau}^*\left(\kappa_2\right)
\nonumber \\
& = & 
g_{\lambda\rho} g_{\sigma\tau}
{\cal{M}}_i^{\lambda\sigma}
{{\cal{M}}_j^{\rho\tau}}^*
= 
{\cal{M}}_i^{\lambda\sigma}
{{\cal{M}}_{\lambda\sigma,j}}^* \,\,\,\,\left(i,j \in
\left\{1,2,3\right\}\right).
\end{eqnarray}
The result for $\mid \!{\overline{\cal{M}}}\! \mid^2$ is too complicated 
to be displayed in this contribution. 
It depends on all Lorentz scalar products which can be constructed from the 
four final-state momentum vectors $k_1$, $k_2$, $q_1$ and $q_2$.

For the neutral decay mode $\eta \rightarrow \pi^0\pi^0\gamma\gamma$ we 
obtain the interactions
\begin{eqnarray}
{\cal{L}}^{(4),2\gamma1\phi}_{WZW} & = & - \frac{e^2 N_c}{24 
\pi^2} \varepsilon^{\mu\nu\alpha\beta} \partial_{\mu} A_{\nu} 
A_{\alpha} \left(
\frac{1}{F_{\pi}}\partial_{\beta}\pi^0
+
\frac{1}{\sqrt{3} F_8}\partial_{\beta}\eta_8
+
\frac{\sqrt{2}}{3 \sqrt{3} \,{\overline{F}}_0}\partial_{\beta}\eta_0
\right),\\
{\cal{L}}^{(2),4\phi} & = & 
-\frac{ F_{\pi}^2 B_0}{24}
\left(
\frac{4 \left(m_u-m_d\right)}{\sqrt{3}F_{\pi}^3 F_8}\eta_8\pi^0\pi^0\pi^0
+
\frac{2 \left(m_u+m_d\right)}{F_{\pi}^2 F_8^2}
\eta_8\eta_8\pi^0\pi^0 \right.\nonumber\\
&&
+
\frac{4 \sqrt{2}\left(m_u+m_d\right)}{F_{\pi}^2 F_8 {\overline{F}}_0}
\eta_8\eta_0\pi^0\pi^0
%\right. \nonumber \\ & & \left.
+
\frac{4 \sqrt{2} \left(m_u-m_d\right)}{\sqrt{3}F_{\pi}^3 {\overline{F}}_0}
\eta_0\pi^0\pi^0\pi^0\nonumber\\
&&\left.
+
\frac{4 \sqrt{2}\left(m_u+m_d\right)}{F_{\pi}^2 F_8 {\overline{F}}_0}
\eta_8\eta_0\pi^0\pi^0
+
\frac{4 \left(m_u+m_d\right)}{F_{\pi}^2 {\overline{F}}_0^2}
\eta_0\eta_0\pi_0\pi_0
\right),
\label{masseq}
\end{eqnarray}
which lead to the diagrams in Fig.~\ref{feyndiagneutral} and 
result in the Lorentz-invariant matrix element
\begin{eqnarray}
{\cal{M}} & = & 
\frac{i e^2 N_c B_0}{96\pi^2} 
\varepsilon^{\mu\nu\alpha\beta} \varepsilon_{1,\nu}
\varepsilon_{2,\alpha}q_{1,\mu}q_{2,\beta}
\left\{
\frac{8\left(m_u-m_d\right)}{\sqrt{3} \left( 
2 q_1 \cdot q_2 - m_{\pi_3}^2
\right)}
\left(
\frac{\cos \theta}{F_{\pi}^2 F_8}
-\frac{\sqrt{2}\sin \theta}{F_{\pi}^2 {\overline{F}}_0}
\right)
\right.\nonumber\\
& & +
\frac{8 \left(m_u+m_d\right)}{3\sqrt{3} \left(
2 q_1 \cdot q_2 - m_{\eta_8}^2
\right)}
\left(
\frac{\cos \theta}{F_8^3}
-\frac{\sqrt{2}\sin \theta}{F_8^2{\overline{F}}_0}
\right)
\nonumber\\
&&\left. +
\frac{32\left(m_u+m_d\right)}{3\sqrt{3}
\left(
2 q_1 \cdot q_2 - m_{\eta_0}^2
\right)
}
\left(
\frac{\cos \theta}{F_8 {\overline{F}}_0^2}
-\frac{\sqrt{2}\sin \theta}{{\overline{F}}_0^3}
\right)
\right\}.
\nonumber\\
&&
\end{eqnarray}
Summing over the possible photon polarizations $\kappa_1$ and $\kappa_2$ 
we obtain a compact result 
for the invariant matrix element squared,
\begin{eqnarray}
\mid\!{\overline{\cal{M}}}\!\mid^2 & = & 
\frac{e^4 N_c^2 B_0^2}{72 
\pi^4}
\left( q_1 \cdot q_2 \right)^2  
\left\{
\frac{m_u-m_d}{\sqrt{3} \left(2 q_1 \cdot q_2 - m_{\pi_3}^2\right)}
\left(
\frac{\cos \theta}{F_{\pi}^2 F_8}
-\frac{\sqrt{2}\sin \theta}{F_{\pi}^2 {\overline{F}}_0}
\right)\right. 
\nonumber\\
&& + 
\frac{m_u+m_d}{3\sqrt{3}\left( 2 q_1 \cdot q_2 - m_{\eta_8}^2\right)}
\left(
\frac{\cos \theta}{F_8^3}
-\frac{\sqrt{2}\sin \theta}{F_8^2{\overline{F}}_0}
\right) 
\nonumber\\
&&\left.
+\frac{4\left(m_u+m_d\right)}{3\sqrt{3}\left( 2 q_1 \cdot q_2 - m_{\eta_0}^2
\right)}
\left(
\frac{\cos \theta}{F_8 {\overline{F}}_0^2}
-\frac{\sqrt{2}\sin \theta}{{\overline{F}}_0^3}
\right)
\right\}^2.
\end{eqnarray}
We note that the 
final result only depends on one Lorentz scalar product, namely $q_1 \cdot 
q_2 = s_{\gamma}/2$. 
\narrowtext

\section{Results and Discussion}

Having determined $\mid\!{\overline{\cal{M}}}\!\mid^2$ we proceed to
investigate the decay spectra by integrating Eq.\ (\ref{width}) numerically.
First we will discuss the Dalitz plot ${\mathrm{d}}^2
\Gamma/{\mathrm{d}}s_{\pi}
{\mathrm{d}}s_{\gamma}$ for the charged decay mode. 
We note that the soft-photon limit for any of the two photons implies 
$s_{\gamma} \rightarrow 0$.
The contour plot for the
full chiral perturbation theory 
calculation (Fig.~\ref{dalfull}) clearly shows the infrared bremsstrahlung
singularity for $s_{\gamma} \rightarrow 0$. 
The Dalitz plot also reflects a pole at $s_{\gamma} = m_{\pi^0}^2 = 
18219 \, {\mathrm{MeV}}^2$ due to the
class 1 Feynman diagram with a propagating $\pi^0$ 
(see Fig.~\ref{feyndiagcharged}). 
In the neighborhood of this pole it is impossible to distinguish between the
processes $\eta\rightarrow\pi^+\pi^-\gamma\gamma$ and $\eta\rightarrow\pi^+
\pi^-\pi^0$. If we switch off the diagrams of class 1, 
the invariant amplitude is
still gauge invariant, and the corresponding Dalitz plot 
(Fig.~\ref{dalwithoutpole}) is very similar to the 
full calculation (Fig.~\ref{dalfull}) except 
for the region around the $\pi^0$ pole. 
We conclude that the diagrams of class 1 do not contribute significantly to the
process which also becomes evident from the diphoton spectrum 
${\mathrm{d}} \Gamma/{\mathrm{d}}z$ (Fig.~\ref{dipion}), where 
$z = s_{\gamma}/m_{\eta}^2$. 
The effect of the $\pi^0$ pole is confined to a very small region around 
$s_{\gamma}=m_{\pi^0}^2$. 
The calculation without class 1 diagrams almost coincides with the full 
calculation except for this region. 
The reaction mechanisms of class 2 
and class 3 diagrams dominate the spectrum over a wide energy range. 
We conclude that the detection of this decay mode should be a good indication 
for the presence of a Wess-Zumino-Witten contact term (class 2). 
Towards small values of $s_{\gamma}$ the bremsstrahlung diagrams
(class 3) will be responsible for a divergence in the spectrum. A measurement
of the steep rise in the spectrum for vanishing $s_{\gamma}$ depends on
the resolution of the detector facilities or, more accurately speaking, on the
minimum photon energy detectable. Hence, in both experiment and theory it
is only possible to determine the partial decay rate as a function of an 
energy cut $\delta m_{brems}$ applied around $\sqrt{s_{\gamma}} = 0$. 
We obtained $\Gamma\left(\eta\rightarrow\pi^+\pi^-\gamma\gamma\right)$ by 
integrating the diphoton spectrum for the calculation without class 1 
diagrams. 
According to Fig.~\ref{dipion} the error introduced by this approximation
should be negligible. 
In Fig.~\ref{partialwidthcharged} we show the result as a
function of $\delta m_{brems}$. 
Comparing our absolute numbers for the partial decay rate with the total eta 
decay rate $\Gamma_{tot} = 1.2 \times 10^{-3}\, {\mathrm{MeV}}$ \cite{PDG94}, 
we note that the resulting branching ratio is well within the reach of 
the facilities mentioned in the introduction.
Using an energy cut at $\delta m_{brems} = 195\,{\mathrm{MeV}}$,
our result for the partial decay rate is smaller than the experimental
upper limits in Eqs.\ (\ref{e1}) and (\ref{e2}) by about three
orders of magnitude.

Let us now turn to the neutral decay mode 
$\eta \rightarrow \pi^0\pi^0\gamma\gamma$.
In contrast to the charged decay mode, where the diagrams of class 1 can be
neglected, these diagrams generate the only tree-level contributions in the
case of neutral mesons. 
For this reason we want to analyse the structure of the vertices
in the diagrams more closely. 
Whereas the two-photon-one-meson vertices have been investigated
extensively in the two-photon decays of the $\pi^0$, $\eta$ and $\eta'$, the
four-meson vertices, in particular the $\eta_8\eta_8\pi^0\pi^0$ interaction,
are not known with comparable precision. 
However, these vertices are of great
interest, because they are directly related to the sum or difference of the
light quark masses (see Eq.\ (\ref{masseq})). While the $\eta\pi\pi\pi$ and 
$\eta'\eta\pi\pi$ interaction can be directly investigated in the decays 
$\eta\rightarrow\pi\pi\pi$ and $\eta'\rightarrow\eta\pi\pi$, this is not the
case for the $\eta\eta\pi\pi$ interaction. Moreover, the minimum center-of-mass
energy for the scattering process
$\pi\pi\rightarrow\eta\eta$ is beyond the convergence radius of chiral
perturbation theory, and $\pi\eta\rightarrow\pi\eta$ scattering is not
a realistic alternative either. 
As a consequence, the only possibility to get information on the 
$\eta\eta\pi\pi$ interaction is the investigation of a composite process 
with at least one additional vertex. 
At first glance the decay $\eta\rightarrow\pi^0\pi^0\gamma\gamma$ 
seems to be a good candidate since $G$ parity is conserved
at the $\eta\eta\pi^0\pi^0$ vertex, whereas in the diagram with a propagating
$\pi^0$ the corresponding vertex violates $G$ parity and vanishes in the
isospin limit. 
However, as is well-known from the decay $\eta\rightarrow\pi\pi\pi$, 
$G$ parity is violated significantly and isospin breaking is reflected by the 
fact that the $\eta\pi\pi\pi$ vertex function is directly proportional to the 
quark mass difference (see Eqs. (\ref{vpvc}) and (\ref{masseq})). 
For this reason the diagram with the propagating neutral pion cannot
be neglected. 
Since $m_{\pi^0}^2$ lies within the range of integration for $s_{\gamma}$, 
the diagram with a propagating $\pi^0$ will cause a pole in the amplitude 
${\cal{M}}$. 
Thus, despite $G$ parity and isospin violation, this diagram is strongly 
enhanced in comparison with the diagrams with a propagating $\eta_8$ and 
$\eta_0$. 
The diphoton spectrum clearly demonstrates the fact 
that the $\eta_8\eta_8\pi^0\pi^0$ interaction plays a minor role in the 
process (see Fig.~\ref{diphoton00}).

In the pole region it is impossible to distinguish
the decay mode $\eta\rightarrow \pi^0\pi^0\gamma\gamma$ from $\eta \rightarrow
\pi^0\pi^0\pi^0$. 
Thus, in order to obtain new information on chiral dynamics from the decay 
$\eta\rightarrow \pi^0\pi^0\gamma\gamma$, one should choose an energy regime 
for $s_{\gamma}$ which is sufficiently far away from this pole.

At ${\cal O}(p^2)$ effective chiral Lagrangians do not reproduce 
the experimentally determined interaction strength of the 
$\eta\pi^0\pi^0\pi^0$ and $\eta'\eta\pi^0\pi^0$ vertices 
(for the correct treatment beyond tree level see \cite{GLe}).
Therefore, we have also performed a phenomenological calculation, 
where we fixed 
these interactions using the experimentally observed branching ratios
\cite{PDG94} 
$\Gamma\left(\eta\rightarrow\pi^0\pi^0\pi^0\right)/\Gamma_{tot} = 0.319$ and 
$\Gamma\left(\eta'\rightarrow\eta\pi^0\pi^0\right)/
\Gamma_{tot} = 0.208$. 
Whereas experimental data agree very well with a {\em constant} tree-level 
prediction for $\eta \rightarrow \pi^0\pi^0\pi^0$,
such an assumption does not seem to work so well for 
$\eta' \rightarrow \eta\pi^0\pi^0$ 
(see, e.g., the discussion in \cite{OW}). 
However, this is of only minor importance in our case, because the pole at 
$m_{\eta'}^2$ in the amplitude is far outside the physical region of 
$s_{\gamma}$. 
It turns out that the result with the vertices fixed by experiment is larger 
than the prediction of chiral perturbation theory by about one order of 
magnitude. 
Furthermore, we also found that $\pi$-$\eta$ and $\pi$-$\eta'$ mixing
according to \cite{bagchi} is negligible in this process. Future work
will focus on higher-order corrections to our tree-level calculation.

The question whether it is realistic to measure the decay 
$\eta\rightarrow\pi^0\pi^0\gamma\gamma$ must be decided from the partial decay
rate $\Gamma\left(\eta\rightarrow\pi^0\pi^0\gamma\gamma\right)$, which we
obtain by integration of the diphoton spectrum applying a cut of the width 
$2\delta m$ around the pole at $m_{\pi^0}$ (Fig.~\ref{neutraltot}). 
From this figure it becomes obvious that this partial decay rate is rather 
small, but possibly within the reach of the new eta facilities. 
However, it will probably be impossible to draw any precise 
information on the $G$-parity-conserving 
$\eta^8\eta^8\pi^0\pi^0$ coupling from measuring this
process, because the corresponding Feynman diagram gives only a small
contribution to the complete amplitude. Finally, we want to
mention that the calculation in \cite{ST} based on old experimental data 
yields
$\Gamma\left(\eta\rightarrow\pi^0\pi^0\gamma\gamma\right)=
1.38\times10^{-3}\,{\mathrm{eV}}$.

We conclude that the rare eta decay $\eta\rightarrow\pi^+\pi^-\gamma\gamma$ is
an interesting test of the anomalous Lagrangian, because it is a simple way to
access the three-meson-two-photon vertex predicted by \cite{WZ} and 
\cite{Witten} and allows for a consistency check with the results from
$\gamma\gamma\rightarrow\pi^+\pi^-\pi^0$. 
The neutral decay mode also investigated in this contribution will be much 
harder to detect.

\newpage
\clearpage

\begin{figure}[h]
\centerline{
%\hspace{4cm}
\epsfxsize=11.5cm
\epsfbox{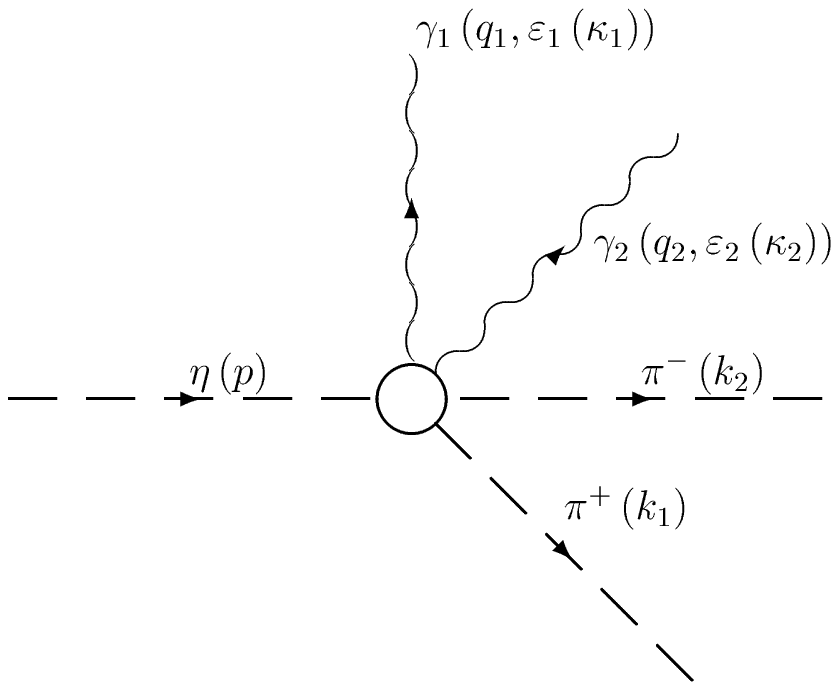}
}
\vspace{0.5cm}
\caption{
\label{kinematics}
\small
Notation for the kinematics of the 
process $\eta \rightarrow \pi^+\pi^-\gamma\gamma$.
}
\end{figure}
\newpage
\clearpage

\begin{figure}[h]
\centerline{
\epsfxsize=8cm
\epsfbox{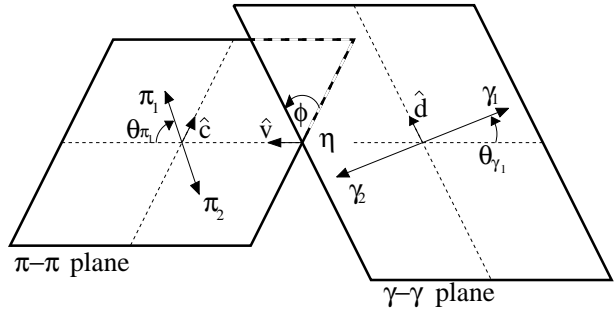}
}
\vspace{0.5cm}
\caption{
\label{planes}
\small
Choice of kinematical variables in the four-particle decay $\eta \rightarrow
\pi\pi\gamma\gamma$.}
\end{figure}

\newpage
\clearpage

\begin{figure}[h]
\centerline{
\hspace{4.5cm}
\epsfxsize=7cm
\epsfbox{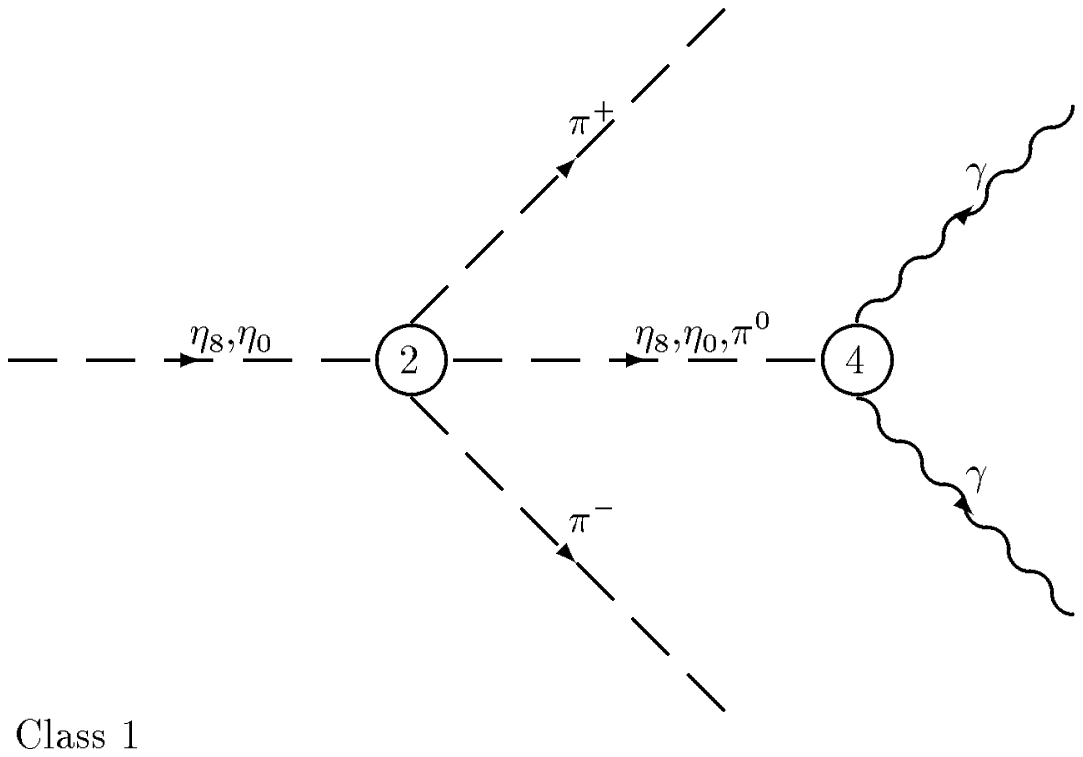}
\hspace{0.5cm}
\epsfxsize=7cm
\epsfbox{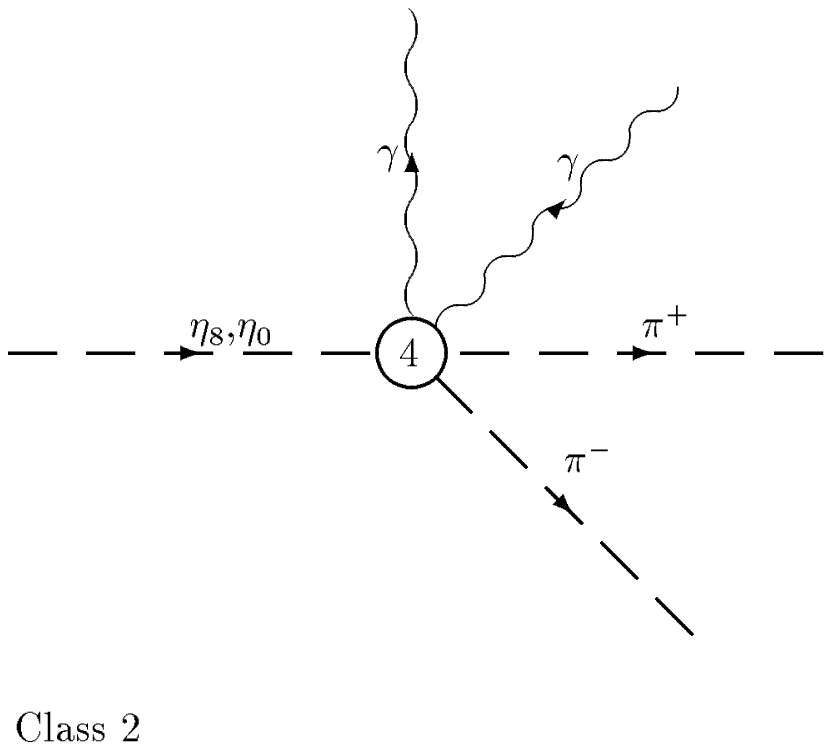}
}
\centerline{
\hspace{4.5cm}
\epsfxsize=7cm
\epsfbox{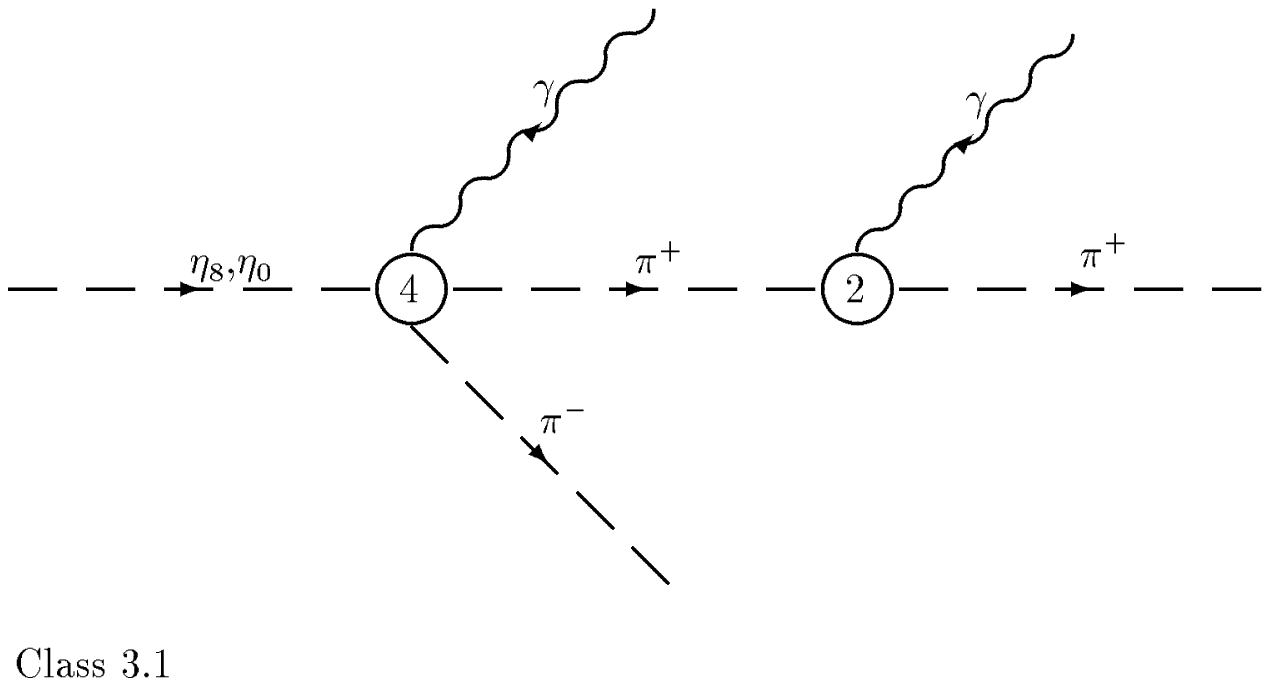}
\hspace{0.5cm}
\epsfxsize=7cm
\epsfbox{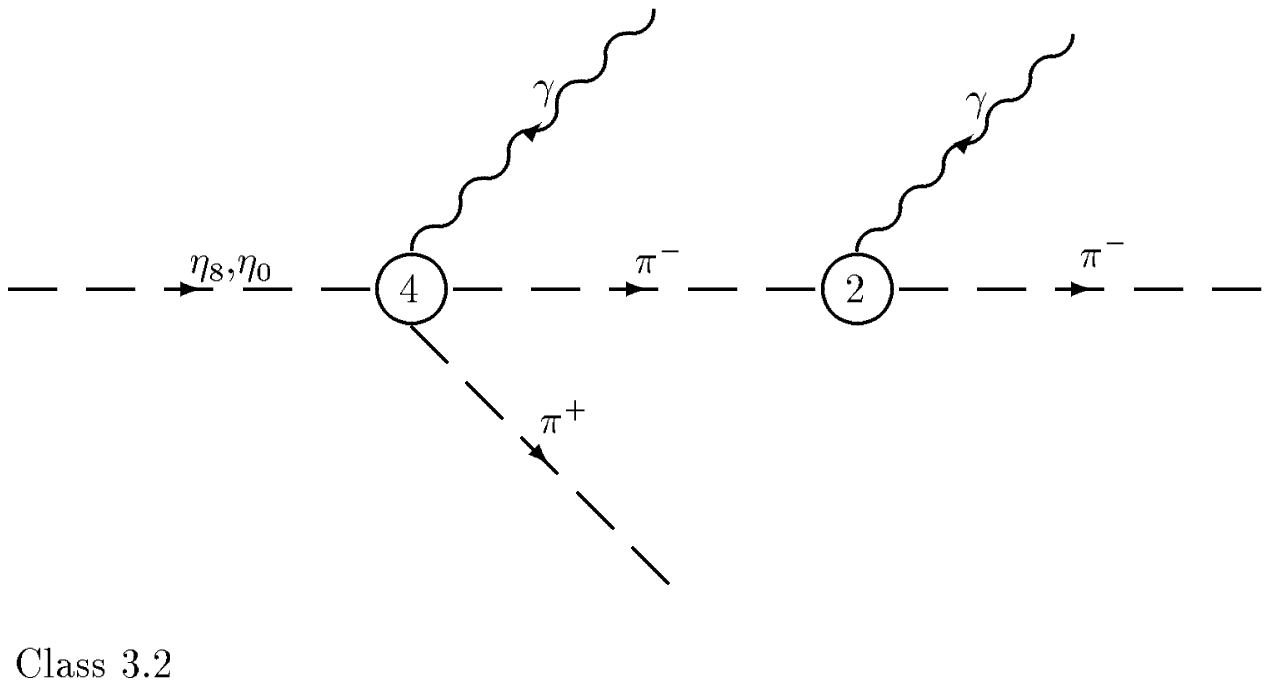}
}
\vspace{0.5cm}
\caption{
\label{feyndiagcharged}
\small
Feynman diagrams of the 
process $\eta \rightarrow \pi^+\pi^-\gamma\gamma$. The numbers in the
interaction blobs denote the order of the vertex in the momentum and quark mass
expansion.
}
\end{figure}

\newpage
\clearpage

\begin{figure}[h]
\centerline{
%\hspace{4cm}
\epsfxsize=9cm
\epsfbox{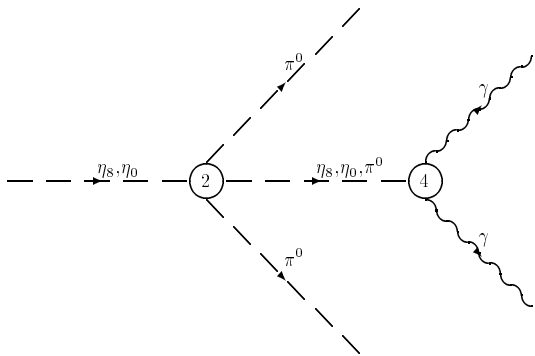}
}
\vspace{0.5cm}
\caption{
\label{feyndiagneutral}
\small
Feynman diagrams of the 
process $\eta \rightarrow \pi^0\pi^0\gamma\gamma$.
}
\end{figure}

\newpage
\clearpage

\begin{figure}[h]
\centerline{
%\hspace{4cm}
\epsfxsize=6.2cm
\epsfbox{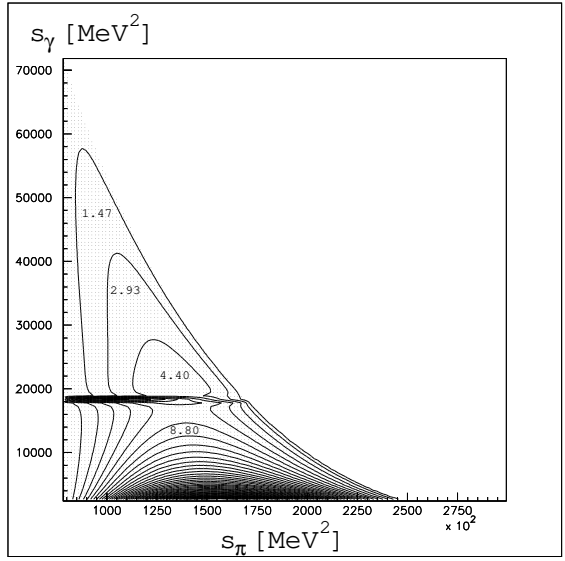}
}
\vspace{0.5cm}
\caption{
\label{dalfull}
\small
Dalitz contour plot for the full calculation in chiral perturbation theory. 
The distance between two contour lines corresponds to a difference of 
$1.47 \times 10^{-17} \, {\mathrm{MeV}}^{-3}$ 
in the doubly differential decay rate 
${\mathrm{d}}^2\Gamma/{\mathrm{d}}s_{\pi}{\mathrm{d}}s_{\gamma}$. The 
outermost contour line denotes the level 
${\mathrm{d}}^2\Gamma/{\mathrm{d}}s_{\pi}{\mathrm{d}}s_{\gamma} = 
1.467 \times 10^{-17} \, {\mathrm{MeV}}^{-3}.$ The dotted area is the 
physical region of the process. The dots in the 
$s_{\pi}$-$s_{\gamma}$ plane represent the points where the 
invariant amplitude was evaluated.
}
\end{figure}

\newpage
\clearpage

\begin{figure}[h]
\centerline{
%\hspace{4cm}
\epsfxsize=6.2cm
\epsfbox{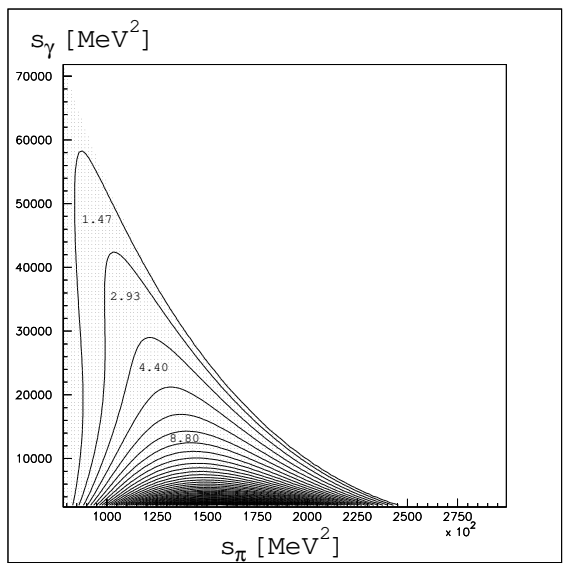}
}
\vspace{0.5cm}
\caption{
\label{dalwithoutpole}
\small
Dalitz contour plot for the 
chiral perturbation theory calculation without class 1 pole diagrams.
The contour spacing is the same as in 
Fig.~\protect{\ref{dalfull}}.
}
\end{figure}

\newpage
\clearpage

\begin{figure}[h]
\centerline{
%\hspace{4cm}
\epsfxsize=8cm
\epsfbox{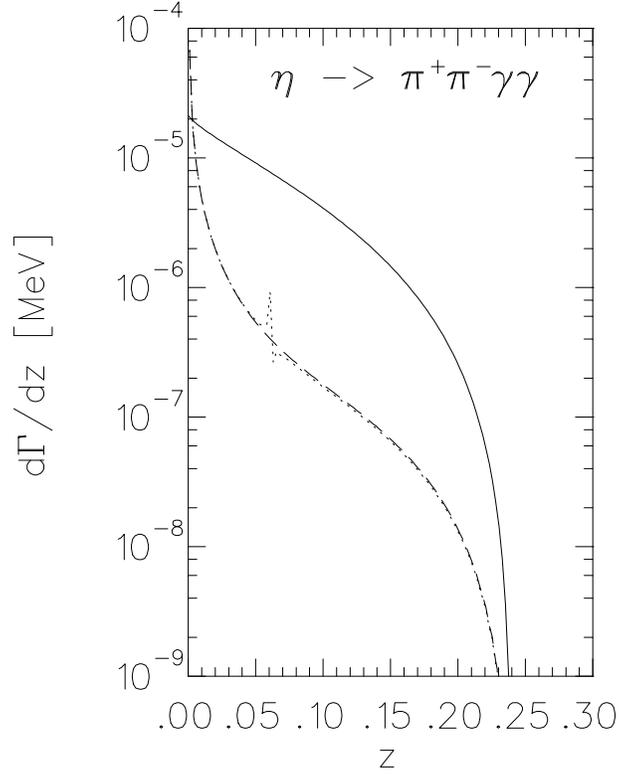}
}
\vspace{0.5cm}
\caption{
\label{dipion}
\small
Diphoton energy spectrum ${\mathrm{d}} \Gamma/{\mathrm{d}}z \,\,\,
(z=s_{\gamma}/m_{\eta}^2)$  
for the 
decay $\eta\rightarrow\pi^+\pi^-\gamma\gamma$. The dotted 
line is the full 
calculation, the dashed line is calculated without class 1 diagrams 
and the solid line is proportional to the phase space integral.
}
\end{figure}

\newpage
\clearpage

\begin{figure}[h]
\centerline{
%\hspace{4cm}
\epsfxsize=8cm
\epsfbox{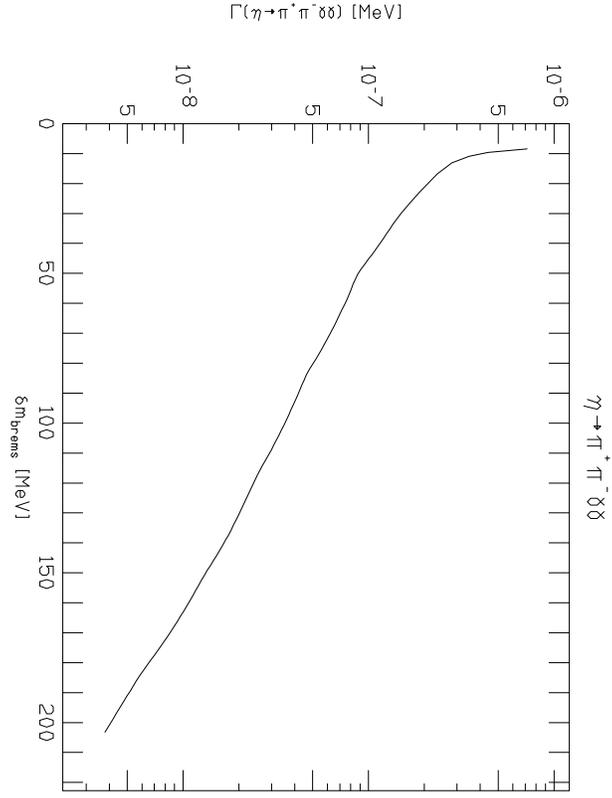}
}
\vspace{0.5cm}
\caption{
\label{partialwidthcharged}
\small
Partial decay rate $\Gamma(\eta\rightarrow\pi^+\pi^-\gamma\gamma)$ 
as a function of the 
energy cut $\delta m_{brems}$ around the bremsstrahlung singularity at 
$s_{\gamma}=0$. 
}
\end{figure}

\newpage
\clearpage

\begin{figure}[h]
\centerline{
%\hspace{4cm}
\epsfxsize=8cm
\epsfbox{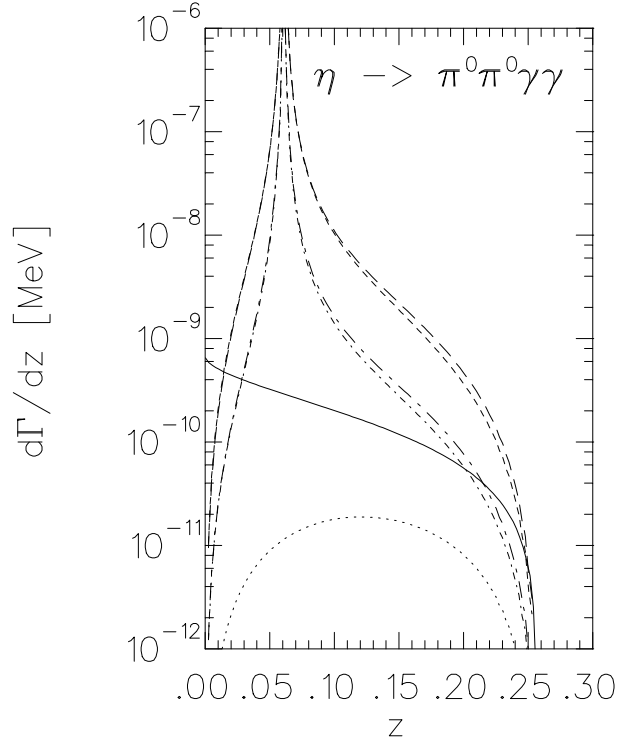}
}
\vspace{0.5cm}
\caption{
\label{diphoton00}
\small
Diphoton spectrum for the decay $\eta \rightarrow \pi^0\pi^0\gamma\gamma$ 
($z=s_{\gamma}/m_{\eta}^2$). The long dash-dotted line is the prediction of
chiral perturbation theory, the short dash-dotted line is the calculation in
chiral perturbation theory, where the $\pi^0\pi^0\eta_8\eta_8$ interaction is
set to $0$, and 
the dotted line is the prediction of chiral perturbation theory for
$m_u = m_d$. The short dashed line is a calculation, where we determined the
strength of the $\eta \pi^0 \pi^0 \pi^0$ interaction from the decay 
$\eta \rightarrow \pi^0\pi^0\pi^0$ and the $\eta\eta'\pi^0\pi^0$ interaction
from the decay $\eta'\rightarrow\eta\pi^0\pi^0$. In the long dashed curve an
additional $\eta_8\eta_8\pi^0\pi^0$ interaction is included. The solid line is
proportional to the phase space integral.}
\end{figure}

\newpage
\clearpage

\begin{figure}[h]
\centerline{
%\hspace{4cm}
\epsfxsize=8cm
\epsfbox{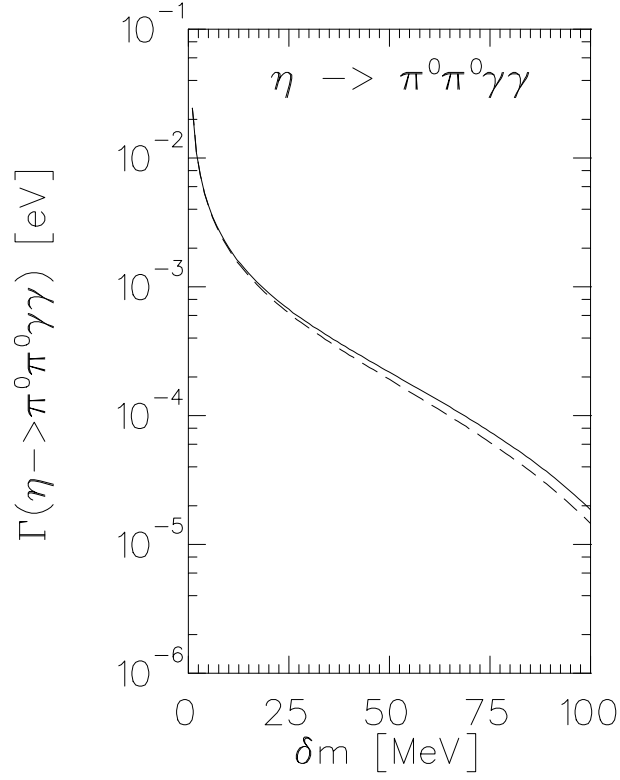}
}
\vspace{0.5cm}
\caption{\label{neutraltot}\small Partial decay
rate for the decay $\eta \rightarrow \pi^0\pi^0\gamma\gamma$ as
a function of the energy cut $\delta m$ in 
$s_{\gamma}^{1/2}$ 
around \protect{$m_{\pi^0}$}. 
The long dashed line is the calculation with the $\eta\pi^0\pi^0\pi^0$ and 
$\eta'\eta\pi^0\pi^0$ vertices fixed by experiment. For the solid line 
we added
a $\eta_8\eta_8\pi^0\pi^0$ interaction from chiral perturbation 
theory.}
\end{figure}

\end{document}